\documentstyle[11pt,epsf]{article}

\newcommand{\lsim}{\;\rlap{\lower 3.5 pt \hbox{$\mathchar \sim$}} \raise 1pt
 \hbox {$<$}\;}

\begin{document}
\title{\vskip-3cm{\baselineskip14pt
\centerline{\normalsize\hfill MPI/PhT/97--006}
\centerline{\normalsize\hfill hep-ph/9705240}
\centerline{\normalsize\hfill February 1997}
}
\vskip1.5cm
Hadronic Higgs Decay to Order $\alpha_s^4$}
\author{K.G. Chetyrkin\thanks{Permanent address:
Institute for Nuclear Research, Russian Academy of Sciences,
60th October Anniversary Prospect 7a, Moscow 117312, Russia.},
B.A. Kniehl, and M. Steinhauser}
\date{}
\maketitle

\vspace{-2em}

\begin{center}
{\it Max-Planck-Institut f\"ur Physik,
    Werner-Heisenberg-Institut,\\ D-80805 Munich, Germany\\ }
\end{center}

\begin{abstract}
We present in analytic form the three-loop ${\cal O}(\alpha_s^2)$ correction
to the $H\to gg$ partial width of the standard-model Higgs boson with
intermediate mass $M_H\ll2M_t$.
Its knowledge is required because the ${\cal O}(\alpha_s)$ correction is so
sizeable that the theoretical prediction to this order is unlikely to be
reliable.
For $M_H=100$~GeV, the resulting QCD correction factor reads
$1+(215/12)\alpha_s^{(5)}(M_H)/\pi
+150.419\left(\alpha_s^{(5)}(M_H)/\pi\right)^2\approx1+0.66+0.21$.
The new three-loop correction increases the Higgs-boson hadronic width by an
amount of order 1\%.
\end{abstract}

\vspace{2em}

The Higgs boson, $H$, is the missing link of the standard model (SM) of
elementary particle physics.
Its experimental discovery would eventually solve the longstanding puzzle as
to whether nature makes use of the Higgs mechanism of spontaneous symmetry
breaking to generate the particle masses.
So far, direct searches at the CERN Large Electron Positron Collider (LEP)
have only been able to rule out the mass range $M_H\le65.6$~GeV at the 95\%
confidence level (CL) \cite{jan}.
On the other hand, exploiting the sensitivity to the Higgs boson via quantum
loops, a global fit to the latest electroweak precision data predicts
$M_H=149{+148\atop-82}$GeV together with a 95\% CL upper bound at 550~GeV
\cite{bou}.

The coupling of the Higgs boson to a pair of gluons, which is mediated at one
loop by virtual quarks \cite{wil}, plays a crucial r\^ole in Higgs
phenomenology.
The Yukawa couplings of the Higgs boson to the quark lines being proportional
to the respective quark masses, the $ggH$ coupling of the SM is essentially
generated by the top quark alone.
The $ggH$ coupling strength becomes independent of the top-quark mass $M_t$ in 
the limit $M_H\ll 2M_t$.
In fact, in extensions of the SM by new fermion generations, this property may
be exploited by using the $ggH$ coupling as a device to count the number of
high-mass quarks \cite{wil}.
In contrast to the electroweak $\rho$ parameter \cite{vel}, the $ggH$ coupling
is also sensitive to quark isodoublets if they are mass-degenerate.
At this point, we also wish to remind the reader that, by the Landau-Yang 
theorem \cite{lan}, spin-one particles such as the photon or the $Z$ boson
cannot couple to two real gluons, while spin-zero particles such as the Higgs
boson do.

The prospects for the Higgs-boson discovery at the CERN Large Hadron Collider
(LHC) vastly rely on the gluon-fusion subprocess, $gg\to H$, which will be the
very dominant production mechanism over the full $M_H$ range allowed
\cite{geo}.
The cross section of inclusive Higgs-boson production in proton-proton
collisions, $pp\to H+X$, is significantly increased, by approximately 70\%
under LHC conditions, by including its leading-order (two-loop) QCD
corrections \cite{daw,djo}, which are intimately related to the $ggH$
coupling.
Under such circumstances, the theoretical prediction for this extremely 
relevant observable can by no means be considered to be well under control,
and it is an urgent matter to compute the next-to-leading-order QCD
corrections at three loops, since there is no reason to expect them to be
negligible.
Recently, a first step in this direction has been taken by considering the 
resummation of soft-gluon radiation in $pp\to H+X$ \cite{kra}.

An important ingredient in this complex research programme is the
${\cal O}(\alpha_s^2)$ three-loop correction to the $ggH$ coupling.
Typical Feynman diagrams that contribute in this order are those obtained by
attaching two virtual gluons to the primary top-quark triangle.
There are also other classes of diagrams, and they all come in large numbers.
The $ggH$ coupling also appears as a building block in the theoretical
description of the crossed process, $H\to gg$, which contributes to the
hadronic decay width of the Higgs boson.
In the low to intermediate mass range, $M_H\lsim150$~GeV, this decay mode has 
a branching fraction of up to 7\% \cite{gro,zer}.
Observing that a Higgs boson in this mass range almost exclusively decays to
$b\bar b$ pairs, this number may be quickly understood by taking the ratio of
the $H\to gg$ and $H\to b\bar b$ partial widths in the Born approximation,
which gives $(\alpha_sM_H/\pi m_b)^2/27$.

The ${\cal O}(\alpha_s)$ correction to the $H\to gg$ decay width was
originally derived \cite{ina} in the limit $M_H\ll2M_t$ by constructing a
heavy-top-quark effective Lagrangian and subsequently confirmed by a
diagrammatic calculation \cite{djo} and via a low-energy theorem (LET)
\cite{let} in Refs.~\cite{daw,djo}.
This correction consists of two-loop contributions connected with $gg$
production and one-loop contributions due to $ggg$ and $gq\bar q$ final
states, where $q$ stands for the first five quark flavours.
In contrast to the $H\to q\bar q$ decay with subsequent gluon radiation, in 
the $H\to gq\bar q$ diagrams of interest here, the $q\bar q$ pair is created 
through the branching of a virtual gluon, so that these contributions
survive in the limit of vanishing $q$-quark mass.
In fact, if all quark masses, except for $M_t$, are nullified, the hadronic
decay width of the Higgs boson is entirely due to $H\to gg$ and the associated
higher-order processes under consideration here.
Depending on the experimental setup, the heavier quarks $Q=c,b$ may be
detectable with certain efficiencies.
The secondary $Q$ quarks from $H\to g g\to gQ\bar Q$ will typically be much
softer than the primary ones from $H\to Q\bar Q\to g Q\bar Q$, which may serve
as a criterion to distinguish between these two production mechanisms.
Alternatively, one may attempt to subtract the $gQ\bar Q$ contributions from 
the QCD-corrected $H\to gg$ decay width \cite{zer}.
For simplicity, following Refs.~\cite{djo,ina}, we shall not consider such a
subtraction for the time being.
Futhermore, as in Refs.~\cite{djo,ina}, we shall concentrate on the limit
$M_H\ll2M_t$, which is most relevant phenomenologically.
Although the LEP1 lower bound on $M_H$ \cite{jan} then implies that $n_l=5$
light quark flavours contribute at the renormalization scale $\mu=M_H$, we
shall keep $n_l$ arbitrary.
Thus, the Born result reads
\begin{equation}
\label{hgg}
\Gamma_{\rm Born}(H\to gg)
=\frac{G_FM_H^3}{36\pi\sqrt2}
\left(\frac{\alpha_s^{(n_l)}(\mu)}{\pi}\right)^2,
\end{equation}
where $G_F$ is Fermi's constant.
The ${\cal O}(\alpha_s)$ correction may be included by multiplying
Eq.~(\ref{hgg}) with \cite{djo,ina}
\begin{equation}
\label{kgg}
K=1+\frac{\alpha_s^{(n_l)}(\mu)}{\pi}\left[\frac{95}{4}-\frac{7}{6}n_l
+\left(\frac{11}{2}-\frac{1}{3}n_l\right)\ln\frac{\mu^2}{M_H^2}\right].
\end{equation}
For $\mu=M_H=100$~GeV, this amounts to an increase by about 66\%.
Since such a sizeable correction is unlikely to provide a useful 
approximation, it is indispensable to go to higher orders.

\begin{figure}[ht]
\leavevmode
\begin{center}
\epsfxsize=16cm
\epsffile[70 530 540 710]{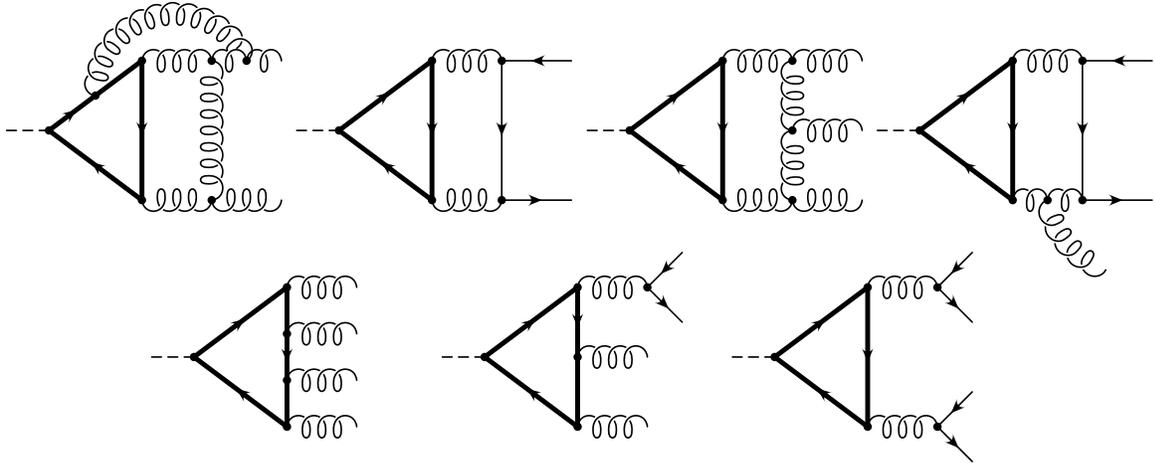}
\caption{Typical diagrams generating ${\cal O}(\alpha_s^2)$
corrections to $\Gamma(H\to gg)$.
Bold-faced (dashed) lines represent the top quark (Higgs boson).}
\label{dia}
\end{center}
\end{figure}

The purpose of this letter is to take the next step by extending
Eq.~(\ref{kgg}) to ${\cal O}(\alpha_s^2)$.
To this end, we need to calculate three-loop three-point, two-loop four-point, 
and one-loop five-point amplitudes.
The contributing final states are $gggg$, $ggq\bar q$,
$q\bar qq^\prime\bar q^\prime$, $ggg$, $gq\bar q$, $gg$, and $q\bar q$.
Typical diagrams are depicted in Fig.~\ref{dia}.

Our procedure is similar to that of Ref.~\cite{ina}.
We construct an effective Lagrangian, ${\cal L}_{\rm eff}$, by integrating out
the top quark.
This Lagrangian is a linear combination of certain dimension-four operators
acting in QCD with five quark flavours, while all $M_t$ dependence is
contained in the coefficient functions.
We then renormalize this Lagrangian and compute with it the $H\to gg$ decay 
width through ${\cal O}(\alpha_s^4)$.
For brevity, we do not list here all operators that enter our analysis in
intermediate steps.
Instead, we immediately proceed to the final version of ${\cal L}_{\rm eff}$,
\begin{equation}
\label{eff}
{\cal L}_{\rm eff}=-2^{1/4}G_F^{1/2}HC_1\left[O_1^\prime\right].
\end{equation}
Here, $\left[O_1^\prime\right]$ is the renormalized counterpart of the bare
operator $O_1^\prime=G_{a\mu\nu}^{0\prime}G_a^{0\prime\mu\nu}$, where
$G_{a\mu\nu}$ is the colour field strength, the superscript 0 denotes bare
fields, and primed objects refer to the five-flavour effective theory.
$C_1$ is the corresponding renormalized coefficient function, which carries
all $M_t$ dependence.
Note that $C_1$ and $\left[O_1^\prime\right]$ are not separately 
renormalization-group (RG) invariant through the order considered, while their
product is.
{}From Eq.~(\ref{eff}) we may derive a general expression for the $H\to gg$ 
decay width,
\begin{equation}
\label{mas}
\Gamma(H\to gg)=\frac{\sqrt2G_F}{M_H}C_1^2
{\rm Im}\,\langle\left[O_1^\prime\right]\left[O_1^\prime\right]\rangle,
\end{equation}
where $\langle\left[O_1^\prime\right]\left[O_1^\prime\right]\rangle$ is 
the vacuum polarization of the Higgs field induced by the gluon operator
at $q^2=M_H^2$, with $q$ being the external four-momentum.

In order to cope with the enormous complexity of the problem at hand, we make
successive use of powerful symbolic manipulation programs.
Specifically, we generate the contributing diagrams with the package QGRAF
\cite{nog} and convert the output to a form that can be used as input for the
packages MINCER \cite{gor} and MATAD \cite{ste}, which solve massless and
massive three-loop integrals, respectively.
The cancellation of the ultraviolet singularities, the gauge-parameter
independence, and the RG invariance serve as strong checks for our
calculation.

We adopt two independent methods to calculate $C_1$.
One is based on the LET \cite{let} and naturally extends the analysis of
Ref.~\cite{ina} by one order in $\alpha_s$.
This leads us to consider the top-quark contributions to the gluon and ghost
propagators as well as the gluon-ghost coupling through ${\cal O}(\alpha_s^4)$
with all external four-momenta put to zero.
Specifically, we need to compute 189, 25, and 228 three-loop diagrams,
respectively.
The external Higgs line is then attached through differentiation with respect 
to the top-quark mass according to the LET.
{}From the resulting three expressions, $C_1$ is then obtained by solving a
linear set of equations \cite{ina}.
The second method is the brute-force calculation of the 657 three-loop 
three-point diagrams which contribute to $C_1$.
Both methods lead to the same result, which upon renormalization reads
\begin{eqnarray}
\label{con}
C_1&=&-\frac{1}{12}\,\frac{\alpha_s^{(6)}(\mu)}{\pi}
\left\{1+\frac{\alpha_s^{(6)}(\mu)}{\pi}
\left(\frac{11}{4}-\frac{1}{6}\ln\frac{\mu^2}{M_t^2}\right)
\right.\nonumber\\
&+&\left.\left(\frac{\alpha_s^{(6)}(\mu)}{\pi}\right)^2
\left[\frac{2693}{288}-\frac{25}{48}\ln\frac{\mu^2}{M_t^2}
+\frac{1}{36}\ln^2\frac{\mu^2}{M_t^2}
+n_l\left(-\frac{67}{96}+\frac{1}{3}\ln\frac{\mu^2}{M_t^2}\right)\right]
\right\},
\end{eqnarray}
where $\alpha_s$ is defined in the $\overline{\rm MS}$ scheme and $M_t$ is the
top-quark pole mass.
Since $C_1$ appears as an overall factor in ${\cal L}_{\rm eff}$, it also
enters the calculation of the $gg\to H$ parton-level cross section at 
next-to-leading order \cite{kra}.
We should mention that Eq.~(\ref{con}) disagrees with the corresponding result
recently found in Ref.~\cite{kra}, although the numerical difference is
relatively small.

In fact, Eq.~(\ref{con}) can be obtained 
from known results  via the following all-order  generalization 
\cite{LET}
of the  LET: 
\begin{equation}
C_1 = - \frac{\pi}{2\alpha_s^{(5)}}
\left(
\frac{%
\displaystyle
\beta^{(6)}(\alpha_s^{(6)})\frac{\partial}{\partial \alpha_s^{(6)}} \alpha_s^{(5)} 
-\beta^{(5)}(\alpha_s^{(5)})
}{1-2\gamma^{(6)}_m(\alpha_s^{(6)})}
\right)
{}.
\label{LET}
\end{equation}
Here 
$\alpha_s^{(6)} =\alpha_s^{(6)}(\mu) $ and $\alpha_s^{(5)}$  
should be understood
as expressed through $\alpha_s^{(6)}(\mu)$ 
via the decoupling relation \cite{lar}
\begin{equation}
\frac{\alpha_s^{(5)}(\mu)}{\alpha_s^{(6)}(\mu)}
 = 
1
-\frac{\alpha^{(6)}_s(\mu)}{\pi}
\left(
\frac{1}{6}l_m
\right)
+\left(\frac{\alpha^{(6)}_s(\mu)}{\pi}\right)^2
\left(
\frac{11}{72}
-\frac{11}{24}l_m
+\frac{1}{36}l^2_m
\right)
+ {\cal O}(\alpha_s^3)  
{},
\label{dec}
\end{equation}
with 
$
l_m = \ln\frac{\mu^2}{m^2_t(\mu)} 
$ 
and $m_t(\mu)$ being the running top quark mass.
Futhermore,  
\begin{eqnarray}
\beta^{(n_f)}(\alpha^{(n_f)}_s) 
&=& - 
\frac{\alpha^{(n_f)}_s}{\pi}
\left\{
\left(11-\frac{2}{3}~n_{{f}}\right)
\frac{\alpha^{(n_f)}_s}{4\pi}
+
\left(102-\frac{38}{3}~n_{{f}} \right)
\left(
\frac{\alpha^{(n_f)}_s}{4\pi}
\right)^2
\nonumber
\right.
\\
&+&
\left.
\left(\frac{2857}{2}-\frac{5033}{18}~n_{{f}} +
\frac{325}{54~}~n_{{f}} ^2\right)
\left(
\frac{\alpha^{(n_f)}_s}{4\pi}
\right)^3
+ {\cal O}(\alpha_s^4)
\right\}
\end{eqnarray}
and 
\begin{equation}
\gamma^{(n_f)}(\alpha_s^{(n_f)}) = 
-\frac{\alpha^{(n_f)}_s}{\pi}
-
\left(\frac{202}{3} - \frac{20}{9}~n_{{f}} \right)
\left(
\frac{\alpha^{(n_f)}_s}{4\pi}
\right)^2
 + {\cal O}(\alpha_s^3)  
\label{gammamass2}
\end{equation}
stand for the $\overline{\rm MS}$ 
beta-function \cite{beta3} and 
the quark anomalous dimension  \cite{gm2},
respectively.
Eq. (\ref{con}) is then recovered from Eq.~(\ref{LET}) by expressing
$m_t(\mu)$ in terms of $M_t$ according to \cite{GraBro90}.

We now turn to the second unknown ingredient in Eq.~(\ref{mas}),
${\rm Im}\,\langle\left[O_1^\prime\right]\left[O_1^\prime\right]\rangle$.
In fact, it is convenient to calculate 
$\langle\left[O_1^\prime\right]\left[O_1^\prime\right]\rangle$ first
and then to take the absorptive part of it.
There is a total of 403 three-loop diagrams to be evaluated.
After renormalization, the result is
\begin{eqnarray}
\label{vac}
{\rm Im}\,\langle\left[O_1^\prime\right]\left[O_1^\prime\right]\rangle
&=&(q^2)^2\frac{2}{\pi}
\left\{1+\frac{\alpha_s^{(n_l)}(\mu)}{\pi}
\left[\frac{73}{4}+\frac{11}{2}\ln\frac{\mu^2}{q^2}
-n_l\left(\frac{7}{6}+\frac{1}{3}\ln\frac{\mu^2}{q^2}\right)\right]
\right.\nonumber\\
&+&\left(\frac{\alpha_s^{(n_l)}(\mu)}{\pi}\right)^2
\left[\frac{37631}{96}-\frac{363}{8}\zeta(2)-\frac{495}{8}\zeta(3)
+\frac{2817}{16}\ln\frac{\mu^2}{q^2}+\frac{363}{16}\ln^2\frac{\mu^2}{q^2}
\right.\nonumber\\
&+&n_l\left(-\frac{7189}{144}+\frac{11}{2}\zeta(2)+\frac{5}{4}\zeta(3)
-\frac{263}{12}\ln\frac{\mu^2}{q^2}-\frac{11}{4}\ln^2\frac{\mu^2}{q^2}
\right)
\nonumber\\
&+&\left.\left.n_l^2\left(\frac{127}{108}-\frac{1}{6}\zeta(2)
+\frac{7}{12}\ln\frac{\mu^2}{q^2}+\frac{1}{12}\ln^2\frac{\mu^2}{q^2}
\right)\right]\right\},
\end{eqnarray}
where $\zeta$ is Riemann's zeta function, with values $\zeta(2)=\pi^2/6$ and
$\zeta(3)\approx1.202$.

We are now in a position to find the ${\cal O}(\alpha_s^2)$ term of the
$K$ factor in Eq.~(\ref{kgg}).
To this end, we insert Eqs.~(\ref{con}) and (\ref{vac}) with $q^2=M_H^2$ into
the master formula~(\ref{mas}) and factor out the Born result of
Eq.~(\ref{hgg}).
In order to get a compact expression, we also eliminate $\alpha_s^{(6)}(\mu)$
in favour of $\alpha_s^{(n_l)}(\mu)$ \cite{lar} and choose $\mu=M_H$.
We thus obtain
\begin{eqnarray}
\label{fin}
K&=&1+
\frac{\alpha_s^{(n_l)}(M_H)}{\pi}\left(\frac{95}{4}-\frac{7}{6}n_l\right)
+\left(\frac{\alpha_s^{(n_l)}(M_H)}{\pi}\right)^2
\left[\frac{149533}{288}-\frac{363}{8}\zeta(2)-\frac{495}{8}\zeta(3)
\right.\nonumber\\
&-&\left.\frac{19}{8}\ln\frac{M_t^2}{M_H^2}
+n_l\left(-\frac{4157}{72}+\frac{11}{2}\zeta(2)+\frac{5}{4}\zeta(3)
-\frac{2}{3}\ln\frac{M_t^2}{M_H^2}\right)
+n_l^2\left(\frac{127}{108}-\frac{1}{6}\zeta(2)\right)\right]
\nonumber\\
&\approx&1+17.917\,\frac{\alpha_s^{(5)}(M_H)}{\pi}
+\left(\frac{\alpha_s^{(5)}(M_H)}{\pi}\right)^2
\left(156.808-5.708\,\ln\frac{M_t^2}{M_H^2}\right),
\end{eqnarray}
where we have substituted $n_l=5$ in the last step.
If we also use the measured values $M_t=175$~GeV and
$\alpha_s^{(5)}(M_Z)=0.118$, and assume $M_H=100$~GeV, we have
\begin{eqnarray}
\label{num}
K&\approx&1+17.917\,\frac{\alpha_s^{(5)}(M_H)}{\pi}
+150.419\,\left(\frac{\alpha_s^{(5)}(M_H)}{\pi}\right)^2
\nonumber\\
&\approx&1+0.66+0.21.
\end{eqnarray}
We observe that the new ${\cal O}(\alpha_s^2)$ term further increases the
well-known ${\cal O}(\alpha_s)$ enhancement by about one third.
If we assume that this trend continues to ${\cal O}(\alpha_s^3)$ and beyond,
then Eq.~(\ref{fin}) may already be regarded as a useful approximation to the
full result.
Inclusion of the new ${\cal O}(\alpha_s^2)$ correction leads to an increase of 
the Higgs-boson hadronic width by an amount of order 1\%.

Equation~(\ref{fin}) may be RG-improved by resumming the terms proportional to
$\ln(M_t^2/M_H^2)$ as described in Ref.~\cite{ina}.
This leads to
\begin{eqnarray}
K&\approx&1+14.938\,\frac{\alpha_s^{(5)}(M_H)}{\pi}
+2.978\,\frac{\alpha_s^{(6)}(M_t)}{\pi}
+104.499\,\left(\frac{\alpha_s^{(5)}(M_H)}{\pi}\right)^2
\nonumber\\
&+&44.491\,\frac{\alpha_s^{(5)}(M_H)}{\pi}\,\frac{\alpha_s^{(6)}(M_t)}{\pi}
+7.818\,\left(\frac{\alpha_s^{(6)}(M_t)}{\pi}\right)^2.
\end{eqnarray}
For the $M_H$ values of interest here (65.6~GeV${}<M_H\ll2M_t$), this amounts
to an insignificant reduction of the absolute value of $K$, by at most 0.6\%,
for $M_H=65.6$~GeV.
In particular, the second line of Eq.~(\ref{num}) remains valid within its
accuracy.

Finally, we wish to mention that the $K$ factor of Eq.~(\ref{fin}) also
applies to the neutral CP-even Higgs bosons of two-Higgs-doublet models
such as the minimal supersymmetric extension of the standard model, as long as
their couplings to gluon pairs are dominantly generated via top-quark loops.

We thank Paolo Nogueira for beneficial communications concerning
Ref.~\cite{nog}.

\end{document}